\documentclass[aps,prd,10pt,onecolumn,nofootinbib,showpacs,showkeys,notitlepage]{revtex4-1}
\usepackage{amsmath,graphicx,bm}
\usepackage{hyperref}

\renewcommand{\d}{\partial}

\newcommand{\nn}{\nonumber\\}

\newcommand{\ph}{\varphi}
\newcommand{\rh}{\varrho}

\newcommand{\exv}[1]{\left\langle{#1}\right\rangle}
\newcommand{\exvs}[1]{\langle{#1}\rangle}
\newcommand{\ep}{\varepsilon}

\renewcommand{\k}{{\bf k}}
\newcommand{\p}{{\bf p}}

\newcommand{\sgn}{\mathop{\textrm{sgn}}}
\newcommand{\Disc}{\mathop{\textrm{Disc}}}

\newcommand{\pint}[2]{{\int\!\frac{d^{#1}#2}{(2\pi)^#1}\,}}
\newcommand{\pintz}[1]{{\int\!\frac{d #1}{2\pi}\,}}

\newlength{\szovszel}\newlength{\szovmag}

\newcommand\lsim{\mathrel{\rlap{\lower4pt\hbox{\hskip1pt$\sim$}} \raise1pt\hbox{$<$}}}                
\newcommand\gsim{\mathrel{\rlap{\lower4pt\hbox{\hskip1pt$\sim$}} \raise1pt\hbox{$>$}}}                

\newcommand{\tehat}{{\ensuremath\quad\Rightarrow\quad}}

\newcommand{\K}{{\cal K}}

\begin{document}

\title{Representation of spectral functions and thermodynamics}

\author{A. Jakov\'ac}
\email{jakovac@phy.bme.hu}
\affiliation{Institute of Physics, Budapest University of Technology
  and Economics, H-1111 Budapest, Hungary}


\date{\today}

\begin{abstract}
  In this paper we study the question of effective field assignment to
  measured or nonperturbatively calculated spectral functions. The
  straightforward procedure is to approximate it by a sum of
  independent Breit-Wigner resonances, and assign an independent field
  to each of these resonances. The problem with this idea is that it
  introduces new conserved quantities in the free model (the new
  particle numbers), therefore it changes the symmetry of the
  system. We avoid this inconsistency by representing each quantum
  channel with a single effective field, no matter how complicated the
  spectral function is.  Thermodynamical characterization of the
  system will be computed with this representation method, and its
  relation to the independent resonance approximation will be
  discussed.
\end{abstract}

\maketitle

\section{Introduction}
\label{sec:intro}

A fundamental field theory is defined through its microscopic field
content, representing the fundamental degrees of freedom (dof), and
the interaction between them. But microscopic dof are not always
observable: in low energy QCD we find hadrons instead of quarks and
gluons. Also in an ordinary gas we might encounter rather molecules
than separate electrons and nuclei (even nucleons). Although this
statement reflects a trivial fact, it is hard to consistently
implement the rather smooth shift/overlap in the nature of the
relevant dof in an interacting theory. If we want to work directly
with the observable dof,
we need to build effective theories.

A primary principle of construction of the effective theories is that
the symmetries of the fundamental system should be present unchanged
in the effective model. This ensures that they belong to the same
universality class, ie. they describe the same physics. In fact, the
by-now textbook examples of the QCD effective models like the
nonlinear or linear sigma models, chiral sigma models, Polyakov loop
models, Walecka model, etc, use this strict guiding principle to
represent low-energy forces generated by QCD \cite{Kapusta:2006pm,
  bwlee}. In these models one defines the field content to represent
the low-lying bound states of QCD, and construct a Lagrangian which
respects the symmetries of QCD. The parameters of the Lagrangian are
eventually tuned to reproduce most observed features of the particle
spectra.

In the examples listed above effective fields represent not just the
(lowest lying) observable hadrons with the corresponding quantum
numbers but actually the complete quantum channel. We get into
trouble, however, when we want to represent individually also the
higher energy states in a given channel; in case of QCD the
plethora of excited hadrons. The representation of the observed
particles would require to assign a field to each hadron. The
manifestation of this idea, treated in the ideal gas approximation is
the widely used hadron resonance gas (HRG) model, which indeed
provides surprisingly accurate description to certain quantities in
QCD in the low temperature regime
\cite{Andronic:2003,Karschetal,Huovinen:2009yb,Borsanyi:2010cj}. For a
more realistic description, in particular to describe transport
properties, one has to consider also the widths of the peaks
(off-shell transport) \cite{Ivanov:1998nv,Ivanov:1999tj}. The problem
of conservation of energy-momentum tensor and other conserved currents
can be consistently solved in the framework of Kadanoff-Baym equations
\cite{Knoll:2001jx}. In this extended version one still assigns a
field (Breit-Wigner resonance) to each peak of the measured spectral
function separately \cite{Knoll:2005nj}.

In these approaches, however, one gives up the principle of keeping
the symmetries of the fundamental theory, by generating a large number
of new conserved quantities at tree level, namely the particle numbers
of the excited hadrons. With other words: each quantum channel is
represented multiple times. If we do not want this unphysical extra
symmetry to appear on the level of observables, the tree level
contributions must be compensated by the effect of interactions. This
may require, however, strong radiative corrections of ${\cal O}(1)$,
eventually resulting in a nonperturbative system. Then tree level
results cannot be considered reliable, whatsoever nice is the arising
description. This situation would question the effective nature of the
chosen model.

There are cases when the idea that we assign a field (a new dof) to
each peak of the spectral functions indeed leads to false
conclusions. A well-known example is the hydrogen atom: if we took all
bound states into account as independent dof then we would obtain
false thermodynamics \cite{Peierls, Landau}. In the standard solution
\cite{Peierls, Landau} we quit at some scale the framework of quantum
mechanics, and explicitly assume that bound states with certain size
cannot exist in the matter. Although physically motivated, nonetheless
this requirement is not self-consistent, and it is hard to imagine
that it can be phrased formally on the level of the Lagrangian. In a
similar spirit, one can try to improve HRG by cutting off the number
of hadronic resonances above an appropriately chosen energy scale
\cite{Cleymans:2011fx}.

We can study this question within the scattering theory, too. In the
quantum mechanical approach, bound states appear as a jump in the
scattering phase. Their contribution to free energy is described by
Beth-Uhlenbeck formula \cite{Landau} provided the scattering phases of
resonances are simply additive \cite{indep}. In the generic situation,
however, this is not the case. Finite width peaks should be included
in the S-matrix as Breit-Wigner resonances with complex amplitudes
\cite{Scon, Svec}. The unitarity of the S-matrix then yields a
consistency constraint, which should be solved. The resulting
amplitudes are thus results of interference effects which might
decrease the ``weight'' of the individual finite lifetime dof. If the
finite width peaks of the spectral function are {\it well separated}, then a
careful analysis can show \cite{RD} that the additivity of their
scattering phases is indeed true.

In this paper we approach this question from the field theory point of
view. The potential problem coming from the introduction of new
conserved quantities can be avoided by simply representing each
quantum channel by a single field -- no matter, how many peaks are in
the corresponding spectral function. As we have discussed above,
associating a single field with the singled out quantum channel works
well in a lot of effective models of QCD. The representation of a
complicated spectral function by a single field, technically is
similar to the perturbative 2PI technique \cite{2PI}, where the
propagator (e.g. the spectral function) of a channel is a single
dynamical object. However, in this case we go beyond the
reorganization of the perturbative series, we shall construct a
non-perturbative scheme where the spectral function is the input. We
expect (and actually will prove) that our scheme shares with the 2PI
technique a number of good features like energy-momentum conservation,
causality, unitarity and renormalizability.

In this paper we try to use this idea to build up a consistent one-dof
bosonic effective model. We start with a very simple analysis,
demonstrating that the correct representation of the spectral function
is relevant (Section \ref{sec:why}). In Section \ref{sec:model} we
present the one-dof model and prove its consistency. In Section
\ref{sec:effects} we calculate the thermodynamical observables (energy
density, pressure etc.) from this model, and compare the results with
the ones assuming independent excitations. The comparison is
facilitated by a prescription how to measure the ``number of dof''. We
will see that overlapping dof are indeed not independent, and they
always behave as if they were less than one degree of freedom. In
Section \ref{sec:conc} we give the conclusions.

\section{Why the representation of the spectral function is
  important?}
\label{sec:why}

The goal of this Section is to demonstrate that the spectral function
alone does not determine the physics completely: the symmetry
properties influence certain observables, in particular the
thermodynamics crucially.

First let us define the spectral function: we choose an operator $A$
with fixed quantum numbers, and compute the $\rh_{AA^\dagger}(t) =
\exvs{[A(t),A^\dagger(0)]}$ expectation value, and take its Fourier
transform $\rh_{AA^\dagger}(p_0)$. If the system is spatially
translation invariant then the spatial momentum is part of the quantum
numbers, too. As it is well known \cite{LeBellac}, this expectation
value at zero temperature is proportional to the density of states, at
finite dimensional quantum mechanical systems it consists of discrete
Dirac-deltas at the position of the energy levels. Choosing different
operators to measure the spectrum results in different weights of the
Dirac-deltas, but their position will be the same.

Now consider a one-dof free scalar Klein-Gordon model. Its Lagrangian
reads
\begin{equation}
  \label{eq:KG}
  {\cal L} = \frac12 \ph K \ph,\qquad K = -\d_t^2 - \omega_k^2,\quad
  \omega_k^2=\k^2+m^2.
\end{equation}
We can use the operator $\ph(t,\k)$ to measure the spectral function
in the momentum $\k$, spin 0 quantum channel. It reads $\rh(p_0)= 2\pi
\sgn(p_0) \delta(p_0^2-\omega_k^2)$. We remark that because of the
canonical commutation relations this spectral function satisfies a sum
rule \cite{LeBellac}.

As a second example we take two scalar degrees of freedom $\ph_1$
and $\ph_2$ with the same mass, and construct their free model:
\begin{equation}
   {\cal L} = \frac12
   \begin{array}[c]{l}(\ph_1,\,\ph_2)\cr\cr \end{array}
   \left(\begin{array}[c]{cc}
     K&  0\cr 0&K\cr \end{array}\right)
   \left(\begin{array}[c]{l}\ph_1\cr\ph_2\cr \end{array}\right).
\end{equation}
To measure the spectral function in the momentum $\k$, spin 0 quantum
channel, we can choose for example the operator
$A(t,\k)=\ph_1(t,\k)\cos\theta+\ph_2(t,\k)\sin\theta$ for arbitrary
angles. The measured spectral function will be, independently on the
angle, $\rh_{AA^\dagger}(p_0)= 2\pi \sgn(p_0)\delta(p_0^2-\omega_k^2)$,
the same result as before.

In this model there is an extra internal symmetry, the O$(2)\equiv$
U$(1)$ rotation of the two fields. If this symmetry is physical, then
there may exist a measurable quantity which makes a distinction
between the two fields; sometimes it is the electric charge attached
to the U$(1)$ rotation. But if we introduce the two fields exclusively
to represent the spectrum, the extra symmetry is non-physical, and the
physical operators do not see the difference between the two degrees
of freedom. In this case physically there exists only one quantum
channel with $\k$ momentum and spin 0.

Now we have two models which give the same spectral function. The
thermodynamics, described by these models, is still different. For
example the Stefan-Boltzmann (SB) limit for the energy density would
be $\ep= \pi^2T^4/30$ in the first case and twice as much in the
second case. This can be physical, of course, if the U$(1)$ symmetry
has a physical background. But if the two degrees of freedom were
introduced exclusively to represent the spectrum, then the extra
symmetry is nonphysical, and we have false results for thermodynamics,
despite the fact that we correctly described the spectrum.

For practical reasons we may still want to choose a two-particle
representation even if the physics do not support the new
symmetry. For example if we have two Dirac-delta peaks in the spectral
function then, according to \cite{RD}, we may use a two-particle
representation. In the limiting case when the two peaks merge, we
arrive to the second model. Since the nonphysical symmetries appear in
the observables, in particular in thermodynamical quantities, to a
correct two-field description we have to choose an interaction which
breaks the artificial symmetry. Now we consider two examples for this
symmetry breaking interaction within the framework of a quadratic
model, where the interaction appears as off-diagonal elements of the
kernel matrix.

Consider for example the following model:
\begin{equation}
   {\cal L} = \frac12
   \begin{array}[c]{l}(\ph_1,\,\ph_2)\cr\cr \end{array}
   \left(\begin{array}[c]{cc}
     K&  K\cr K&K\cr \end{array}\right)
   \left(\begin{array}[c]{l}\ph_1\cr\ph_2\cr \end{array}\right)
\end{equation}
This Lagrangian has no more U$(1)$ symmetry. A special property of
this model is that the kernel has no inverse. This means that one
degree of freedom (now the $\ph_1-\ph_2$ mode) has no dynamics, it
completely decouples from the system. The remaining mode $\ph_1+\ph_2$
has the same dynamics as in \eqref{eq:KG}, so the physics is the same
too. In particular the spectral function (up to a normalization
factor\footnote{We can determine the spectral function measured by the
  the above defined $A$, it results in $\rh_{AA}(p_0)
  =\pi\sin^2(\theta+\frac\pi4)\sgn(p_0)\delta(p_0^2-\omega_k^2)$. This
  depends on $\theta$ since the action is not U$(1)$ symmetric}) and
the thermodynamics is the same.

Another example is the model
\begin{equation}
   {\cal L} = \frac12
   \begin{array}[c]{l}(\ph_1,\,\ph_2)\cr\cr \end{array}
   \left(\begin{array}[c]{cc}
     K-\lambda\quad &  -\lambda \cr -\lambda\quad 
     &K-\lambda\cr \end{array}\right)
   \left(\begin{array}[c]{l}\ph_1\cr\ph_2\cr \end{array}\right).
\end{equation}
The two eigenvalues are $K$ and $K-2\lambda$, so the spectral function
exhibits two peaks. But in case $\lambda\to\infty$, the mass of one
eigenstate remains $m$, while the other has a mass-squared
$m^2+2\lambda\to\infty$ which decouples from the system. So the
measurable part of the spectral function as well as the thermodynamics
at any finite temperature behave as if we had one dof.

So apparently we can work with more dof than the physical ones. The
price we have to pay for it is that we have strong interactions. This
must be so, since in the non-interacting (diagonal) case the SB limit
is twice as much as in the interacting case: so the interactions give
${\cal O}(1)$ corrections relative to the free case. The lesson is
that wrong assignment of the dof's alone may lead to an apparently
strongly interacting theory, even if with correctly chosen fields the
theory is free.

\section{Representation of a generic spectral function with a single
  field}
\label{sec:model}

Let us try to generalize the observations of the previous Section for
the case when we observe, after fixing all possible quantum numbers, a
generic spectral function. Let us assume that we can identify finite
width peaks. Heuristically, the width of the peaks can be treated as a
``resolution power''. If, according to this resolution, the separation
of the peaks is large, then earlier studies \cite{RD,Svec} tell us
that these behave as distinct objects: the independent particle
approximation works well. If the separation of the peaks starts to be
comparable with their width, then by their own resolution they are
more and more like one object, and we arrive at the case discussed
above. Then the lesson of the examples of Section~\ref{sec:why} tell
us that if we further insist to a multi-dof description, the theory
which correctly describes physics becomes strongly interacting,
nonperturbative.

This complication can be avoided if we do not introduce new dof into
the system, and assign a single field to each quantum channel: then
the symmetries of the original system is not altered. As a price,
however, we have to work with a more complicated quadratic theory
which can represent a generic spectral function. In this Section we
construct a 1-dof system which can represent a generic spectral
function, then we demonstrate that it describes a consistent theory,
if the input spectral function comes from a physical system.

\subsection{The Lagrangian}
\label{sec:lag}

Let us assume that we have measured a scalar spectral function, and
there are no other hidden quantum numbers in the system. This means
that the system must be represented with a single scalar field $\ph$.
We are looking for a quadratic Lagrangian which could represent the
given spectral function (interactions can be necessary to correctly
describe spectral functions of other quantum channels, which is not
the topic of the present paper). We want moreover a system where
energy and momentum are conserved, so we want space and time
translation symmetry. Then the most generic scalar field theory can be
written as:
\begin{equation}
  \label{eq:S0}
  S[\ph] = \frac12 \int d^4x\,\ph(x) \K(i\d)\,\ph(x).
\end{equation}
In Fourier space we can write
\begin{equation}
  S[\ph] = \frac12 \pint4p \ph^*(p) \K(p) \ph(p).
\end{equation}
If $\K$ is not a polynomial, then this action is not local. Since
$\ph(x)$ is real, $\ph(-p)=\ph^*(p)$; since $S$ is also real, the
kernel must satisfy $\K^*(p)=\K(-p)=\K(p)$.

The spectral function is defined as
\begin{equation}
  \rh(x-y)=\exv{[\ph(x),\ph(y)]}.
\end{equation}
Since the theory is quadratic, so the commutator $[\ph(x),\ph(y)]$ is a
c-number, thus we can also omit the expectation value. Using the
standard procedure \cite{LeBellac}, we compute the retarded retarded
Green's function (Landau prescription) as
\begin{equation}
  \label{GRK}
  G_R(p) = \K^{-1}(p_0+i\ep,\p).
\end{equation}
The spectral function is the discontinuity of the retarded Green's
function
\begin{equation}
  \rh(p) = \Disc_{p_0} G_R(p) = \lim_{\ep\to
    0}\left[\K^{-1}(p_0+i\ep,\p) - \K^{-1}(p_0-i\ep,\p)\right].
\end{equation}
That means that a kernel determines the spectral function in a unique
way.

In our approach the spectral function is the input, and we want to
construct the corresponding kernel. The retarded Green's function can
be obtained via the Kramers-Kronig relation:
\begin{equation}
  \label{KK}
  G_R(p) = \pintz\omega \frac{\rh(\omega,\p)}{p_0-\omega+i\ep}.
\end{equation}
Using $\eqref{GRK}$, the kernel can be determined as
\begin{equation}
  \label{eq:K}
  \K(p)  = \left(\emph{P}\pintz\omega
    \frac{\rh(\omega,\p)}{p_0-\omega}\right)^{-1},
\end{equation}
where $P$ means principal value. That means that the kernel is also
completely determined by the spectral function.

\subsection{Consistency}

When one works with a nonlocal theory like \eqref{eq:S0}, one has to
worry about the consistency of the model which, in case of a field
theory, means that the requirements of energy- and momentum
conservation, causality, unitarity, and Lorentz-invariance (in case of
zero temperature) must be satisfied.

First we remark that within perturbation theory it is also possible to
change the tree level propagator and work with a self-consistently
determined generic propagator instead: this is the 2PI resummation
\cite{2PI}. This method is known to be a fully consistent way of
treating the system perturbatively. Since one can formulate 2PI
resummation as a model with nonlocal kernel \cite{Jakovac:2006gi}, the
present method can be thought to be a nonperturbative realization of
the consistent 2PI resummation method.

But, independently on the perturbative treatment we can prove the
consistency criteria one-by-one.

\paragraph{Energy and momentum conservation}

This requirement is satisfied if the action is time and space
translation invariant. From the definition \eqref{eq:S0} this fact is
trivial.

\paragraph{Causality:}

Causality means in relativistic invariant systems that measurements at
space-like separated points must not influence each other (in
non-relativistic systems causality is not an issue). Mathematically
this requires that local operators at space-like separated positions
must commute. We show that if the input spectral function is causal
then the system is causal, too.

A local operator built up from a single scalar field has the form
$\hat O(x) = \prod_{i=1}^n A_i$, where $A_i= a_i(x,i\d) \ph(x)$
where $a_i(x,i\d)$ is a local differential operator (ie. at most
polynomial in the derivatives). We define $\hat Q(y)$ in a similar way
as $\prod_{i=1}^m B_i$. Then we can use the following identity
\begin{equation}
  [\hat O(x),\hat Q(y)] = \sum_{i=1}^n \sum_{j=1}^m B_1\dots B_{j-1}
  A_1\dots A_{i-1} [A_i,B_j] A_{i+1}\dots A_n B_{j+1}\dots B_m.
\end{equation}
The commutator can be expressed as $[A_i,B_j]= a_i(x,i\d_x)
b_j(y,i\d_y) [\ph(x),\ph(y)]$. Since $a_i$ and $b_j$ are local
differential operators, an infinitesimally small opened neighborhood
of the point $x$ is enough to compute them from the field $\ph$. But
if $x$ and $y$ are spatially separated points, ie. $(x-y)^2<0$, then a
sufficiently small opened neighborhood is also spatially
separated. Since $\rh(x-y) = [\ph(x),\ph(y)]$ is the input spectral
function, it is guaranteed that it is zero for spatial
separation. This means that in a small neighborhood it is identically
zero, therefore all derivatives are zero. This means that $[A_i,B_j]$
is zero $\forall\,i,j$, and finally $[\hat O(x),\hat Q(y)]=0$, too.

\paragraph{Unitarity}

Physically unitarity means that there is no loss of probability in the
system. If we use, for example, finite lifetime particles, then the
time evolution will not be unitary, since the decaying state will be
missing from the final states. To circumvent this problem we have to
keep all possible energy states, then a decaying state turns into
another state and the complete probability will be the same. As a
dynamical quantity, therefore, we have to work with the complete
spectral function.

Mathematically one can examine the unitarity (and the corresponding
reflection positivity for euclidean systems) of models where the
kernel is an arbitrary polynomial \cite{Polonyi:2010zv}. The result
is, formulating in the language of the spectral function, that for
unitarity the spectral function must be positive for all positive
frequencies $\rh(\omega>0,\p) \ge 0$. Since we use the spectral
function as an input stemming from a real quantum field theory, this
requirement will be satisfied. Therefore unitarity is also granted.

\paragraph{Lorentz invariance}

From the definition \eqref{eq:S0} it is clear that if the kernel is
Lorentz invariant, then the action is Lorentz invariant, too. For the
reverse argumentation, in Lorentz invariant systems the input spectral
function leads to a Lorentz invariant retarded Green's function, which
leads to a Lorentz invariant kernel. So, for a physically sensible
zero temperature spectral function the Lorentz invariance of our model
is also granted.

The input spectral function, if it comes from a finite temperature
system, need not be Lorentz invariant. Still, in most part of this
paper we will use a Lorentz invariant input.

We see, that all requirement of a consistent field theory could be
satisfied with the Ansatz \eqref{eq:S0}, provided we use an input
spectral function which is physically sensible. Now we are in the
position that we can use it for calculating physical observables.

\section{Thermodynamics}
\label{sec:effects}

Given the Lagrangian of the system, in principle we can calculate the
expectation value of any physical observable, and, in particular,
approach thermodynamics. But, because of time-nonlocality, some
standard techniques do not work in this case. First of all we do not
have canonical formalism (more precisely we would need infinitely many
fields to represent the canonical formalism). Therefore there is not a
direct connection between the imaginary time formalism and the
thermodynamics: in fact, the standard derivation is based on the fact
that the Hamiltonian density depends on the canonical momentum as
$\Pi^2/2$ \cite{LeBellac}. Therefore thermodynamics should be started
from something which is meaningful also microscopically: this can be
the energy density. Known for all temperatures, this is then enough to
calculate all the other thermodynamical quantities like pressure or
entropy density.

\subsection{Energy-momentum tensor}
\label{sec:EMtens}

The energy density is the 00 component of the energy-momentum tensor,
which represents the Noether currents belonging to the time and space
translation invariance, respectively. We can use the standard
procedure to determine the energy-momentum tensor: we consider
position dependent translation $a^\mu(x)$ with which ${x'}^\mu =
x^\mu+a^\mu(x)$ and the shifted field satisfies
$\ph'(x')=\ph(x)$. Then we can calculate the action with the shifted
field $S[\ph']$, for which we expect the behavior $S[\ph']-S[\ph] =
\int d^4x\, T^{\mu\nu}(x)\,\d_\mu a_\nu$, where $T^{\mu\nu}$ is the
energy momentum tensor.

\emph{Statement:} The energy momentum tensor belonging to the
action \eqref{eq:S0} is
\begin{equation}
  T_{\mu\nu}(x) = \frac12 \ph(x)\,D_{\mu\nu}\K(i\d)\, \ph(x)
\end{equation}
where
\begin{equation}
  D_{\mu\nu}\K(i\d) = \left[\frac{\d\K(p)} {\d p^\mu} \biggr|_{p\to
      i\d}\right]_{sym}\!\!\! i\d_\nu - g_{\mu\nu}\K(i\d),
\end{equation}
and the symmetrized derivative is defined as
\begin{equation}
  f(x) [(i\d)^n]_{sym} g(x) = \frac1{n+1}\sum\limits_{a=0}^{n} [(-i\d)^a
  f(x) ] [(i\d)^{n-a} g(x)].
\end{equation}

\emph{Proof}: We represent the kernel in Fourier space as a
power series:
\begin{equation}
  \K(p)=\sum c_{(n)}^{\mu_1\dots\mu_n}p_{\mu_1}\dots p_{\mu_n}\tehat
   \K(i\d)=\sum c_{(n)}^{\mu_1\dots\mu_n}(i\d_{\mu_1})\dots (i\d_{\mu_n}),
\end{equation}
where $c$ is a completely symmetric tensor. Now we consider $S[\ph']$
with integration variable $x'$, then change to $x'\to x$ where
$x'=x+a$ with position dependent $a$. The corresponding Jacobian is
${\d x'_\mu}/{\d x^\nu} = \delta_\mu^\nu + \d_\mu a^\nu + {\cal
  O}(a^2)$, and the transformation of the derivative reads $\d'_\mu =
\d_\mu - (\d_\mu a^\nu) \d_\nu + {\cal O}(a^2)$. We obtain
\begin{eqnarray}
  S[\ph'] &&= \sum c_{(n)}^{\mu_1\dots\mu_n} \int d^4x' \ph'(x')
  (i\d'_{\mu_1})\dots (i\d'_{\mu_n}) \ph'(x') =\nn&& = \sum
  c_{(n)}^{\mu_1\dots\mu_n} \int d^4x (1+\d a) \ph(x) \,(i\d_{\mu_1} -
  \d_{\mu_1} a^{\nu_1} i\d_{\nu_1})\dots (i\d_{\mu_n} - \d_{\mu_n}
  a^{\nu_n} i \d_{\nu_n}) \, \ph(x).
\end{eqnarray}
We need the difference linear in $\d a$. We find
\begin{equation}
  \delta S = \int d^4 x \d a {\cal L} - \sum
  c_{(n)}^{\mu_1\dots\mu_n} \int d^4x \ph(x) \sum_{i=1}^n
  (i\d_{\mu_1})\dots (i\d_{\mu_{i-1}}) \d_{\mu_i} a^{\nu_i}
  i\d_{\nu_i} (i\d_{\mu_{i+1}})\dots (i\d_{\mu_n}) \, \ph(x).
\end{equation}
Now we perform partial integration in the second term, and find
\begin{equation}
  - \sum c_{(n)}^{\mu_1\dots\mu_n} \int d^4x \sum_{i=1}^n  \d_{\mu_i} a^{\nu_i}
  \left[(-i\d_{\mu_1})\dots (-i\d_{\mu_{i-1}}) \ph(x)\right]
  \left[(i\d_{\mu_{i+1}})\dots (i\d_{\mu_n}) \,(i\d_{\nu_i}) \ph(x)\right]. 
\end{equation}
Since $c_{(n)}^{\mu_1\dots\mu_n}$ is completely symmetric in the
indexes, we can put $i\to n$ and find
\begin{equation}
  - \sum c_{(n)}^{\mu_1\dots\mu_{n-1}\mu} \int d^4x \d_\mu a^\nu \sum_{i=1}^{n-1}
  \left[(-i\d_{\mu_1})\dots (-i\d_{\mu_{i-1}}) \ph(x)\right]
  \left[(i\d_{\mu_{i}})\dots (i\d_{\mu_{n-1}}) \,(i\d_\nu\ph(x))\right].
\end{equation}
Introducing the symmetrized derivative notion we find
\begin{equation}
  - \sum nc_{(n)}^{\mu_1\dots\mu_{n-1}\mu} \int d^4x \d_\mu a^\nu \ph(x)
  \left[(i\d_{\mu_1})\dots (i\d_{\mu_{n-1}})\right]_{sym}
  \,(i\d_\nu\ph(x)) =  \int d^4x \d_\mu a^\nu \ph(x) \left[\frac12
    \frac{\d\K(p)} {\d p_\mu}\biggr|_{p\to i\d}\right]_{sym}\!\!\!(i\d_\nu\ph(x)).
\end{equation}
Finally
\begin{equation}
  \delta S =  \frac12\int d^4 x  \d^\mu a^\nu(x)\left\{g_{\mu\nu}
    \ph(x)\K(i\d)\ph(x) - \ph(x) \left[\frac{\d\K(p)} {\d p^\mu}
      \biggr|_{p\to i\d}\right]_{sym}\!\!\!(i\d_\nu\ph(x))\right\},
\end{equation}
so we find
\begin{equation}
  T^{\mu\nu}(x) = \frac12 \ph(x) \left\{ \left[\frac{\d\K(p)} {\d p^\mu}
      \biggr|_{p\to i\d}\right]_{sym}\!\!\! i\d_\nu -
    g_{\mu\nu}\K(i\d) \right\}\ph(x).
\end{equation}
This is the equation we had to prove.\\ \emph{QED}

\subsection{Finite temperature expectation value}

Now we will compute the expectation value of the energy-momentum
tensor in equilibrium. For that we change into Fourier space:
\begin{equation}
  T_{\mu\nu}(k) = \frac12 \pint4p\frac{d^4q}{(2\pi)^4}
  (2\pi)^4\delta(k-p-q) \ph(q)\, D_{\mu\nu}\K(q,p)\,\ph(p),
\end{equation}
where the symmetrization of the derivatives is inherited to the
momentum dependent kernel as
\begin{equation}
  \label{eq:pnsym}
  [p^n]_{sym}\to \frac1{n+1}\sum_{i=0}^n (-q)^{i} p^{n-i}.
\end{equation}
We can take its expectation value:
\begin{equation}
  \label{Tmnexp1}
  \exv{T_{\mu\nu}(k)} = \frac12\pint4p\frac{d^4q}{(2\pi)^4}
  (2\pi)^4\delta(k-p-q) D_{\mu\nu}\K(q,p)\,\exv{\ph(q)\ph(p)} =
  \frac12 \pint4p D_{\mu\nu}\K(q=-p,p)\, i G_<(p) (2\pi)^4\delta(k).
\end{equation}
In the expression of $\K(q=-p,p)$, according to \eqref{eq:pnsym}, the
symmetrization of the derivative can be omitted, so we simply denote
\begin{equation}
  \label{eq:Dmunu}
  D_{\mu\nu}\K(p) = p_\mu \frac{\d\K}{\d p^\nu} - g_{\mu\nu}\K.
\end{equation}
Since \eqref{Tmnexp1}  is proportional to $\delta(k)$, in real space it
is position-independent: $\exv{T_{\mu\nu}(x)} = \exv{T_{\mu\nu}}$. We
remark that if the kernel is relativistic invariant (i.e. depends only
on $p^2$) then $\exv{T_{\mu\nu}}$ is a symmetric tensor.

To determine the propagator at finite temperature we use KMS relation
\cite{LeBellac}. We note here that the formal correspondence between
the time translation $e^{-iHt}$ and statistical operator $e^{-\beta
  H}$ is still valid, and so the KMS relation remains true. With the
KMS relation we find:
\begin{equation}
  \label{eq:Tmunuexp}
  \exv{T_{\mu\nu}} = \frac12 \pint4p D_{\mu\nu}\K(p)\, n(p_0)\rh(p),
\end{equation}
where $n(p_0)=(e^{\beta p_0}-1)^{-1}$ is the Bose-Einstein
distribution.  Using the fact that $\K$ is symmetric real function,
$D_\mu \K$ is also symmetric real function, while $\rh(p_0)$ is a real
antisymmetric function. Therefore we can average the positive and
negative energy parts as
\begin{equation}
  n(p_0)\rh(p_0,\p) \to \left(\frac12+n(p_0)\right)\rh(p_0,\p),
\end{equation}
which is a completely symmetric function. We can then restrict
ourselves to the positive frequency part as
\begin{equation}
  \exv{T_{\mu\nu}} = \int\limits_p^+ D_{\mu\nu}\K(p)\,
  \left(\frac12+n(p_0)\right)\rh(p), \quad\mathrm{where}\quad
  \int\limits_p^+ = \int\limits_0^\infty \frac{dp_0}{2\pi} \pint3\p.
\end{equation}
The leading $1/2$ represents the vacuum energy. To obtain a finite
expression we have to renormalize the expression which means ``normal
ordering'', ie. subtract the expectation value at zero temperature:
\begin{equation}
  \exv{T_{\mu\nu}}_{ren} = \int\limits_p^+ D_{\mu\nu}\K(p)\,\left[
    n(p_0)\rh(p) + \frac12 \delta\rh(p)\right],\qquad
  \mathrm{where}\quad \delta\rh(p) = \rh(p)-\rh_0(p).
\end{equation}
In a rotationally invariant system only the diagonal elements
survive. The energy density is the $00$ component of the
energy-momentum tensor:
\begin{equation}
  \label{eps}
  \ep = \int\limits_p^+ D\K(p)\,\left[ n(p_0)\rh(p) +
    \frac12\delta\rh(p)\right], \qquad D\K(p) = p_0 \frac{\d\K}{\d p^0}\, -
  \,\K,
\end{equation}
This is a completely finite expression for the energy density. This
formula is the main result of this paper.

In case of relativistic invariance the kernel is a function of $p^2$
alone, then we can write
\begin{equation}
   D\K(p) =2p_0^2 \frac{\d\K}{\d p^2}\, - \,\K.
\end{equation}

Note, that \eqref{eps} is \emph{non-linear} functional of the input
spectral function, since apart from the explicit $\rh$, the kernel
$\K$ also depends on it. This nonlinearity has the interesting
consequence that the result is independent of the overall
normalization of the spectral function. Indeed, if we change the
normalization of $\rh$, ie. we multiply the original spectral function
by $Z$, then $G_R\to ZG_R$ and $\K\to \K/Z$, and so $Z$ drops out from
the result. This fact can be interpreted that thermodynamics is not
sensitive to the normalization of the fields, only on the density of
states.

Once we know the $\ep(T)$ relation, we can use thermodynamics to
calculate other quantities. For the explicit formulae we will not
consider thermal variation of the spectrum, and so assume $\delta
\rh=0$. In that case the pressure is
\begin{equation}
  \label{pressure}
  p = -T \int\limits_p^+ \frac1{p_0}\,D\K(p)\,\ln\left(1-e^{-\beta
      p_0}\right) \rh(p),
\end{equation}
and the entropy density reads
\begin{equation}
  s=  \int\limits_p^+ \frac1{p_0}\,D\K(p)\,\biggl[ \beta p_0 n(p_0) -
  \ln\left(1-e^{-\beta p_0}\right) \biggr]\rh(p).
\end{equation}
These expressions are also well defined finite quantities.

\subsection{Number of degrees of freedom}

The real observables are the energy density and the corresponding
thermodynamical quantities. In order to help understanding the physics
of the effective system, other, more loosely defined quantities can be
useful like ``number of dof''.

For stable free particles this notion is well defined: it is the
number of energy levels at fixed quantum numbers, in particular at
fixed momentum. The spectral function $\rh(\omega,\k)$ consists of
Dirac-delta peaks in this case, and the number of dof is the
\emph{number} of the Dirac-delta peaks. The heights do not matter,
they come from the normalization of the operator with which we measure
the spectrum. Note however, that the normalization independence means
that the number of dof must be a nonlinear functional of the spectral
function!

A general input spectral function can have arbitrary structure,
usually broadened peaks and other ``continuum'' contributions. The
question of the number of dof is thus not well-defined there. We may
hope, however, that plausible definitions yield approximately the same
physical picture.

A plausible choice is, for example, to compute the energy density $\ep$
of the system. For a free theory with $N_{\mathrm{dof}}$ degrees of
freedom the energy density is $N_{\mathrm{dof}}\ep_0(m,T)$, where
$\ep_0(m,T)$ is the free energy density of a gas with mass $m$ at
temperature $T$. This quantity, therefore, is proportional to the
number of dof, so we can define a ``thermodynamical dof'' at any
temperature. The definition depends on the mass $m$, unless we go to
asymptotic high temperatures, where $N_{\mathrm{dof}}(T\to\infty) =
30\ep(T\to\infty)/(\pi^2T^4)$.

Another proposal is to use the formula of the energy density and use
the ideas of Williams and Weizsacker equivalent photon number
\cite{WW}. The corresponding definition is the ``zero temperature
equivalent dof'' with the Ansatz:
\begin{equation}
  \label{np}
  N_\mathrm{dof} = \int\limits_0^\infty \frac{dp_0}{2\pi}\, \frac1{p_0}
  D\K(p)\,\rh(p),
\end{equation}
This is a dimensionless, temperature independent quantity which, as we
will see later, in case of the discrete spectrum yields indeed the
number of energy levels.

\section{Independence of the degrees of freedom}

We have all the formulae to answer physical questions. The main issue
will be to compare the results coming from the independent resonance
approximation and the one-field representation.

\subsection{Dirac-delta peaks}
\label{sec:Diracpeaks}

The first problem that we should check is whether a system that
contains $N$ stable excitations with different dispersion relations
can be represented with $N$ free fields. According to \cite{RD} we
expect that it must be true.

Consider now a spectral function consisting of $N$ Dirac-delta peaks
with dispersion relations $\omega_i(\p)$ ($\omega_i\neq\omega_j$ for
$i\neq j$), allowing also different normalization:
\begin{equation}
  \label{Ndirdel_rho}
  \rh(p) = \sum\limits_i 2\pi Z_i \delta(p_0-\omega_i(\p)).
\end{equation}
In the independent resonance approximation we assign the following
Lagrangian to this spectral function:
\begin{equation}
  {\cal L}_{indep} = \sum\limits_{i=1}^N \frac12
  \ph_i(i\d_0-\omega_i(i\d)) \ph_i.
\end{equation}
The normalization is irrelevant, since we can rescale the fields to
achieve the standard form. Using this Lagrangian and measuring the
spectral function of the operator $\ph=\sum_i\sqrt{Z_i}\ph_i$ we indeed
recover \eqref{Ndirdel_rho}. 

This Lagrangian represents an $N$-dof system. If we calculate the 
thermodynamical observables, the energy density and the pressure we
obtain sum of the partial energy densities and pressure:
\begin{equation}
  \label{epandp_indep}
  \ep = \sum\limits_{i=1}^N \ep_i,\qquad P = \sum\limits_{i=1}^N P_i,
\end{equation}
where
\begin{equation}
  \label{epiandpi}
  \ep_i = \pint3\p \omega_i n(\omega_i),\qquad 
  P_i = \pint3\p \frac p3 \frac{d\omega_i}{dp} n(\omega_i).
\end{equation}

Let us check whether we obtain the same result in the single-field
representation, where -- correctly -- we do not introduce new
conserved charges into the system. Now we have a single field; from
the Kramers-Kronig relation \eqref{KK} we find for the retarded
Green's function
\begin{equation}
  G_R = \sum\limits_i \frac{Z_i}{p_0-\omega_i+i\ep}.
\end{equation}
The kernel is its inverse: it is now a complicated, nonlocal object
which have zeroes at $p_0=\omega_i(\p)$ dispersion
relations. Therefore we immediately see that $\K\rh=0$. The nonzero
contribution to the energy density \eqref{eps}, pressure
\eqref{pressure} or number of dof \eqref{np} is the derivative of the
kernel multiplied by the spectral function. In case of the energy
density we need
\begin{equation}
  p_0\frac{\d\K}{\d p_0} = -\frac{p_0}{G_R^2} \frac{\d G_R}{\d p_0} =
  \frac{p_0}{G_R^2}  \sum\limits_i \frac{Z_i}{(p_0-\omega_i)^2}.
\end{equation}
The spectral function evaluates this formula at each
$p_0=\omega_i$. We find
\begin{equation}
  \lim_{p_0\to\omega_i} p_0\frac{\d\K}{\d p_0} =  \lim_{p_0\to\omega_i} 
  \frac{\omega_i}{Z_i^2/(p_0-\omega_i)^2+\mathrm{finite}}\left[
    \frac{Z_i}{(p_0-\omega_i)^2 + \mathrm{finite}} \right] =
  \frac{\omega_i}{Z_i}.
\end{equation}
That means
\begin{equation}
  D\K(p)\,\rh(p)= \sum\limits_{i=1}^N
  \lim_{p_0\to\omega_i}\left(p_0\frac{\d\K}{\d p_0}\right) Z_i(2\pi)
  \delta(p_0-\omega_i) =\sum\limits_{i=1}^N \omega_i (2\pi)
  \delta(p_0-\omega_i).
\end{equation}
The wave function renormalization factors disappear, and the energy
becomes $\omega_i$. Therefore the number of degrees of freedom from
\eqref{np} is reduced to the integral
\begin{equation}
  N_\mathrm{dof} = \sum\limits_{i=1}^N \int\limits_0^\infty
  \frac{dp_0}{2\pi}\, \frac1{p_0}\,\omega_i (2\pi)
  \delta(p_0-\omega_i) = N,
\end{equation}
supporting the $N$ independent degrees of freedom
approximation. The formula of the energy density \eqref{eps} yields
\begin{equation}
    \ep =\sum\limits_{i=1}^N \int\limits_p^+ n(p_0)\,\omega_i (2\pi)
    \delta(p_0-\omega_i) =\sum\limits_{i=1}^N \pint3\p \omega_i n(\omega_i),
\end{equation}
which agrees with \eqref{epandp_indep} with
\eqref{epiandpi}. Similarly for the pressure, using \eqref{pressure},
we recover the pressure formula in \eqref{epandp_indep} with
\eqref{epiandpi}.

We can conclude that for a spectral function which has $N$
Delta-peaks, the independent resonance approximation exactly
reproduces the results of the one-field representation. This also
agrees with the earlier results \cite{RD}.

\subsection{Single particle with finite width}

Now let us start to study the problem of overlapping peaks. The basic
building block is the case of a single particle with finite
width. There the spectral function has the (relativistic) Breit-Wigner
form
\begin{equation}
  \rh(p) = \frac{4p_0\Gamma}{(p_0^2-\Gamma^2-\omega_\p^2)^2+4p_0^2\Gamma^2}.
\end{equation}
From this spectral function we obtain the retarded Greens function and
the kernel as
\begin{equation}
  G_R(p) = \frac1{(p_0+i\Gamma)^2-\omega_p^2},\qquad
  \K(p) = p_0^2-\Gamma^2-\omega_p^2,\qquad D\K= p_0^2+\Gamma^2+\omega_p^2.
\end{equation}
The formula for the number of degrees of freedom \eqref{np} can be
exactly evaluated:
\begin{equation}
  N_\mathrm{dof} = \int\limits_0^\infty \frac{dp_0}{2\pi}\,
  \frac{p_0^2+\Gamma^2+\omega_p^2}{p_0}\,
  \frac{4p_0\Gamma}{(p_0^2-\Gamma^2-\omega_\p^2)^2+4p_0^2\Gamma^2} = 1,
\end{equation}
it is independent on the width of the Lorentzian.

\begin{figure}[hbtp]
  \centering
  \includegraphics[height=5cm]{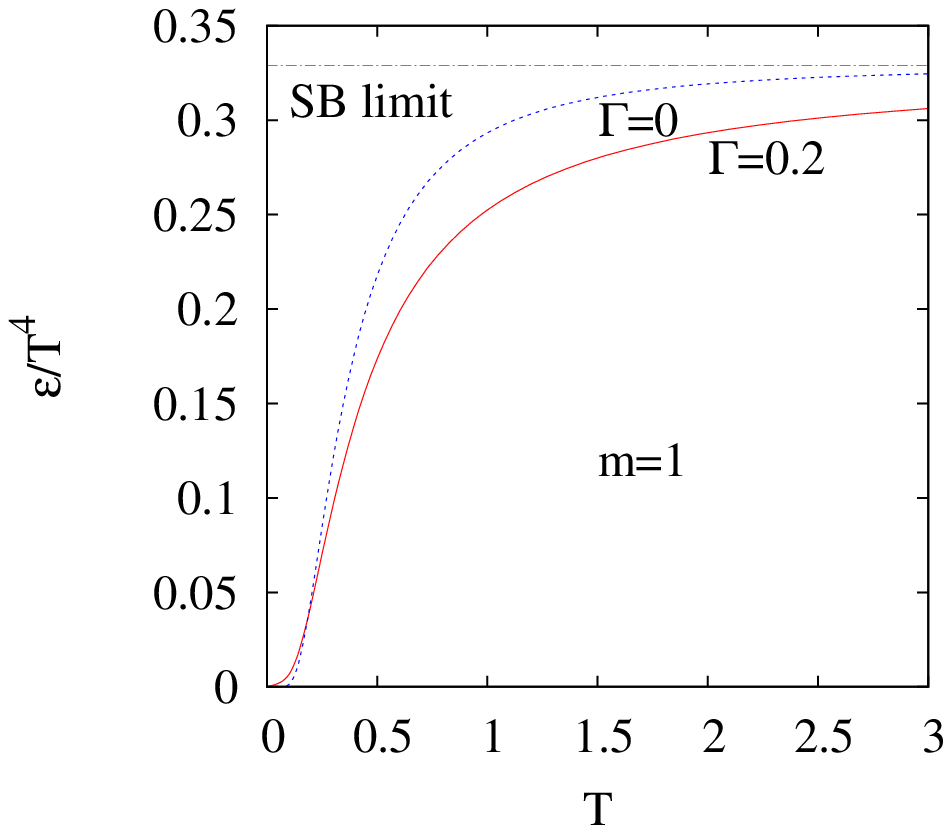}
  \includegraphics[height=5cm]{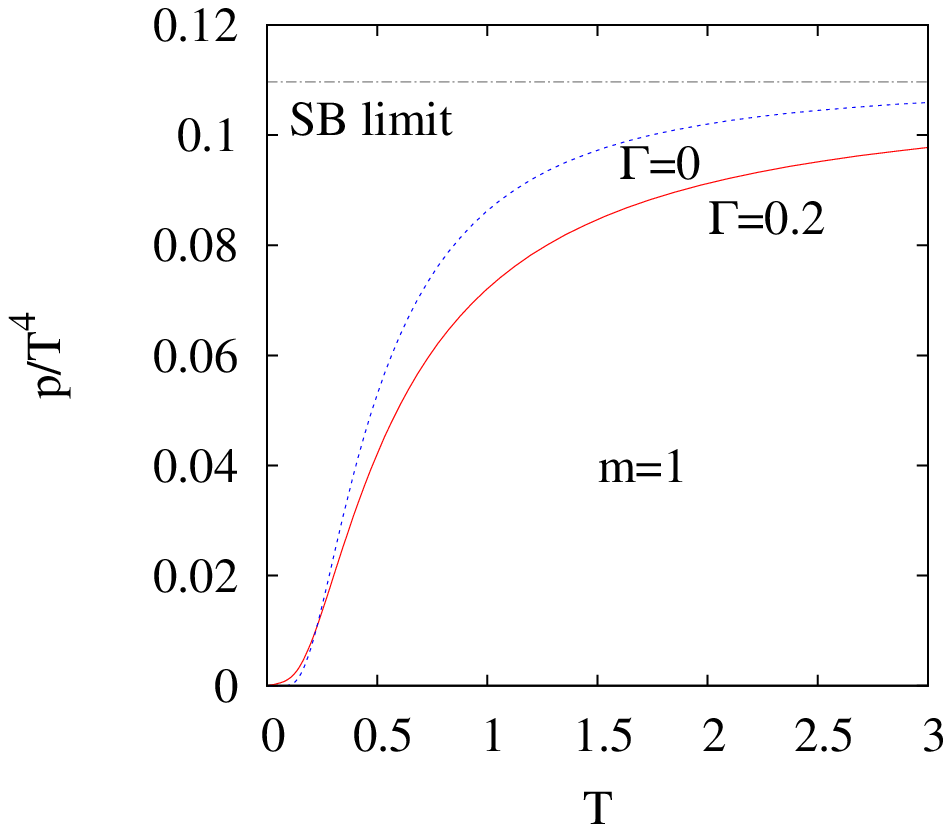}\\
  \hspace*{2em}\emph{a.)}\hspace*{20em}\emph{b.)}
  \caption{The $T^4$ normalized \emph{(a.)} energy density and
    \emph{(b.)} pressure of a system with spectral function of a
    single Lorentzian, compared to the same mass, zero width case. SB
    means Stefan-Boltzmann limit.} 
  \label{fig:Lorentz_energydensity}
\end{figure}
For the energy density and pressure we have to do numerical
calculations. The result can be seen on
Fig. \ref{fig:Lorentz_energydensity}. For comparison we have depicted
the zero width (ie. Dirac-delta) case. We can see that despite the
relatively large with ($\Gamma/m=0.2$), the energy density and
pressure are very close to the free case, especially in the low energy
region. This makes understandable that the hadron resonance gas
approximation works well in the low energy regime even for particles
with relatively large width like the rho-meson.

\subsection{Two Lorentzians}

\begin{figure}[htbp]
  \centering
    \includegraphics[height=5cm]{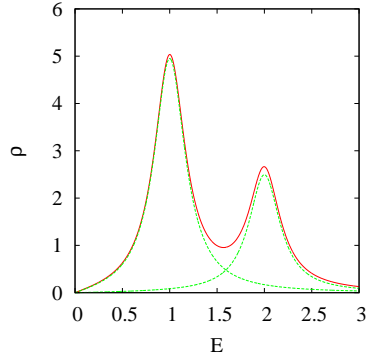}
  \caption{Spectral function consisting two Breit-Wigner form with the
    same width ($\Gamma=0.2)$ and with masses $m=1$ and $2$. The
    dashed lines show the two Breit-Wigner ingredients of the spectral
    function.}
  \label{fig:twolor_rho}
\end{figure}

Now let us take two Lorentzians in the same quantum channel. The
spectral function which we use consists of two independent
Breit-Wigner functions:
\begin{equation}
  \label{twolorrho}
  \rh(p) =
  \frac{4Z_1p_0\Gamma_1}{(p_0^2-\Gamma_1^2-\omega_1^2)^2+4p_0^2\Gamma_1^2} +
  \frac{4Z_2p_0\Gamma_2}{(p_0^2-\Gamma_2^2-\omega_2^2)^2+4p_0^2\Gamma_2^2}. 
\end{equation}
The spectral function is shown on Fig. \ref{fig:twolor_rho}.
Now the spectral function can be \emph{exactly} represented by two
finite width particles. The independent resonance approximation then
would yield the sum of the single Lorentzian results: the number of
degrees of freedom is $2$, and the energy-density and pressure is the
sum of the one-Lorentzian results:
$\ep=\ep_{1-Lorentzian}(m_1)+\ep_{1-Lorentzian}(m_2)$ and
$p=p_{1-Lorentzian}(m_1)+p_{1-Lorentzian}(m_2)$.

Now we work out the result coming from the one-field representation
which does not introduce new conserved quantities. The corresponding
retarded Greens function is:
\begin{equation}
  G_R(p) = \frac{Z_1}{(p_0+i\Gamma_1)^2-\omega_1^2} +
  \frac{Z_2}{(p_0+i\Gamma_2)^2-\omega_2^2},
\end{equation}
its inverse is the kernel of the Lagrangian, according to
\eqref{eq:K}. From there we can calculate the number of dof
\eqref{np}, the energy density \eqref{eps} and pressure
\eqref{pressure}. These letter need numerical calculations, the number
of dof, however, can be worked out analytically, and it reflects well,
what is going on when the two peaks start to merge.

\begin{figure}[htbp]
  \centering
    \includegraphics[height=5cm]{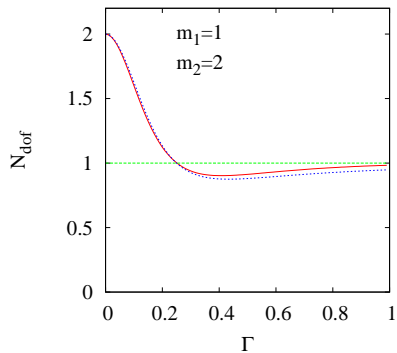}
  \caption{The number of degrees of freedom at fixed masses but
    changing width of the two Lorentzian peaks. The solid line is the
    definition, the dashed line comes from the approximate formula
    \eqref{eq:Npapp}.}
  \label{fig:Ndof}
\end{figure}

By representing the spectral function with its pole structure, the
formula for the number of dof \eqref{np} reads in this case $
N_\mathrm{dof} =  N_\mathrm{dof}^{(1)} + N_\mathrm{dof}^{(2)}$ where
\begin{equation}
  N_\mathrm{dof}^{(1)} =\int\limits_{-\infty}^\infty
  \frac{dp_0}{2\pi}\, \frac{DK(p)}{p_0}\,
  \frac{Z_1}{4i m_1} \left[ \frac1{p_0-i\Gamma_1- m_1}
    -\frac1{p_0-i\Gamma_1+ m_1} 
    -\frac1{p_0+i\Gamma_1- m_1} +\frac1{p_0+i\Gamma_1+ m_1} \right],
\end{equation}
and $ N_\mathrm{dof}^{(2)}$ is a similar expression with
$1\leftrightarrow2$ index change. We can close the contour of the
integration from above (or below, the result is the same) to find
\begin{equation}
  N_\mathrm{dof}^{(1)} = \frac{Z_1}{4 m_1}
  \left[\frac{DK(p)}{p_0}\biggr|_{p_0= m_1+i\Gamma_1}
    - \frac{DK(p)}{p_0}\biggr|_{p_0=- m_1+i\Gamma_1} \right].
\end{equation}
Since $D\K(-p_0)=D\K(p_0)$, the above expression is real. The actual
expression is rather complicated, but with help of Mathematica one can
work out the analytical formulae. The generic formula is rather
lengthy, here we analyze the $Z_1=Z_2$ and $\Gamma_1=\Gamma_2$
case. Then the result reads
\begin{equation}
  N_\mathrm{dof}^{(1)} = \frac{(2048 \Gamma^8 + ( m_1^2 -  m_2^2)^4 + 16
    \Gamma^2 ( m_1^2 -  m_2^2)^2 ( m_1^2 +  m_2^2) +
    512 \Gamma^6 (5  m_1^2 + 3  m_2^2) + 32 \Gamma^4 (23
     m_1^4 + 34  m_1^2  m_2^2 + 7  m_2^4)} {(64
    \Gamma^4 + ( m_1^2 -  m_2^2)^2 + 16 \Gamma^2 (3  m_1^2
    +  m_2^2))^2}
\end{equation}
and for $ N_\mathrm{dof}^{(2)}$ we obtain the same formula with
$1\leftrightarrow2$ change. When we choose $m_1=1,\, m_2=2$ then we
can plot the degrees of freedom, it is shown in Fig. \ref{fig:Ndof}.
To understand better these formulae it is worth to work out the
approximate formula valid in the $\Gamma,\, |m_1- m_2| \ll m_{1,2}$. We
obtain
\begin{equation}
  \label{eq:Npapp}
  N_\mathrm{dof}= 1 + \frac{1-y^2}{(1+y^2)^2},\qquad \mathrm{where}\qquad
  y = \frac{4\Gamma}{|m_2- m_1|}.
\end{equation}
Although it is just an approximate formula, it approximates well the
correct expression, as Fig.  \ref{fig:Ndof} shows.

The robust feature of the number of dof is that for small width it
starts at $2$: this is the case of two Dirac deltas analyzed in
Subsection \ref{sec:Diracpeaks}. On the other hand as
$\Gamma\to\infty$ the number of dof goes to one. The two peaks merge,
and there is no way to say whether they were two separate peaks or
not. Then the situation is similar to the one we considered in Section
\ref{sec:why}. We can therefore trace how a two-degrees-of-freedom
system becomes a one-degree of freedom system dynamically, without
changing the number of representing fields manually.

At non-extremal cases, ie. when $0<\Gamma<\infty$, the number of dof
is not a well-defined physical observable. As we can see, with the
present definition \eqref{np} the number of dof can be smaller than
one. Still, in order to determine the width when the the originally
two-dof system turns into a one-dof system, this definition gives
probably a sensible answer. Therefore we can claim that already at
$\Gamma/\Delta m\approx 0.2$ we have an approximately one-dof system,
although in the spectral function the two peaks are clearly
identifiable yet (cf. Fig.  \ref{fig:twolor_rho}).

We remark here that this phenomenon is closely related to the
resolution of the Gibbs paradox. The paradox is that if in a two dof
system we continuously make the ``difference'' of the dof to
disappear, then for arbitrarily small difference we observe a 2-dof
system which suddenly, non-analytically changes to a 1-dof system at
zero difference. The resolution, according to the above analysis is
that this change is in fact analytical if we consider finite lifetime
particles. A similar conclusion is drawn in \cite{GibbsQM}. Physically
it means that we cannot perform a sufficiently long measurement which
is accurate enough to identify the two separate energy levels in case
of finite lifetime.

\begin{figure}[hbtp]
  \centering
  \includegraphics[height=5cm]{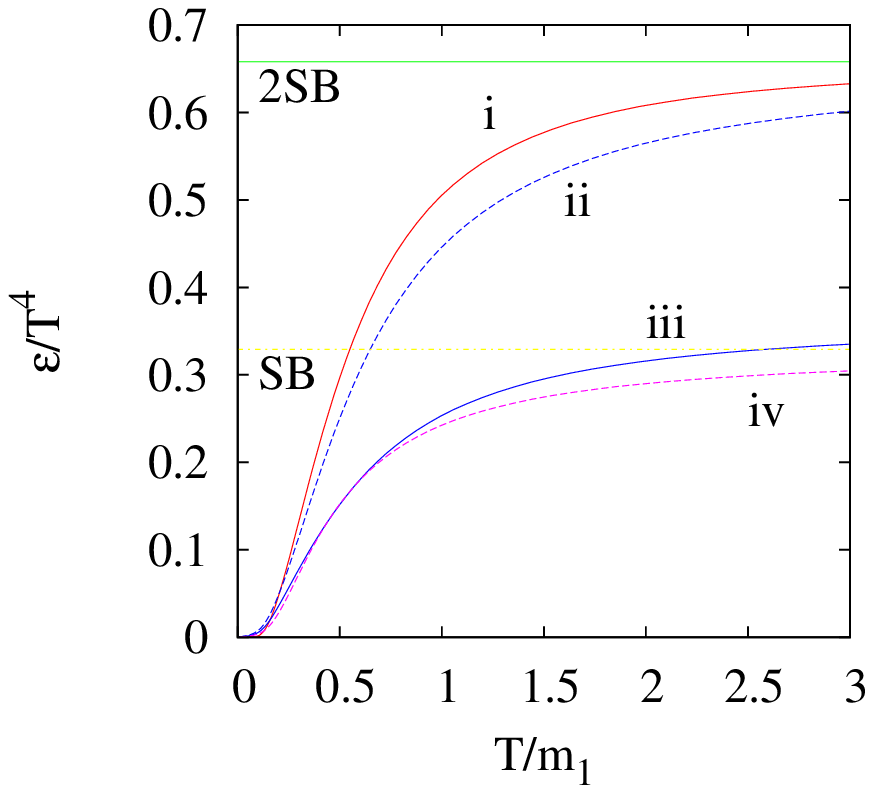}
  \includegraphics[height=5cm]{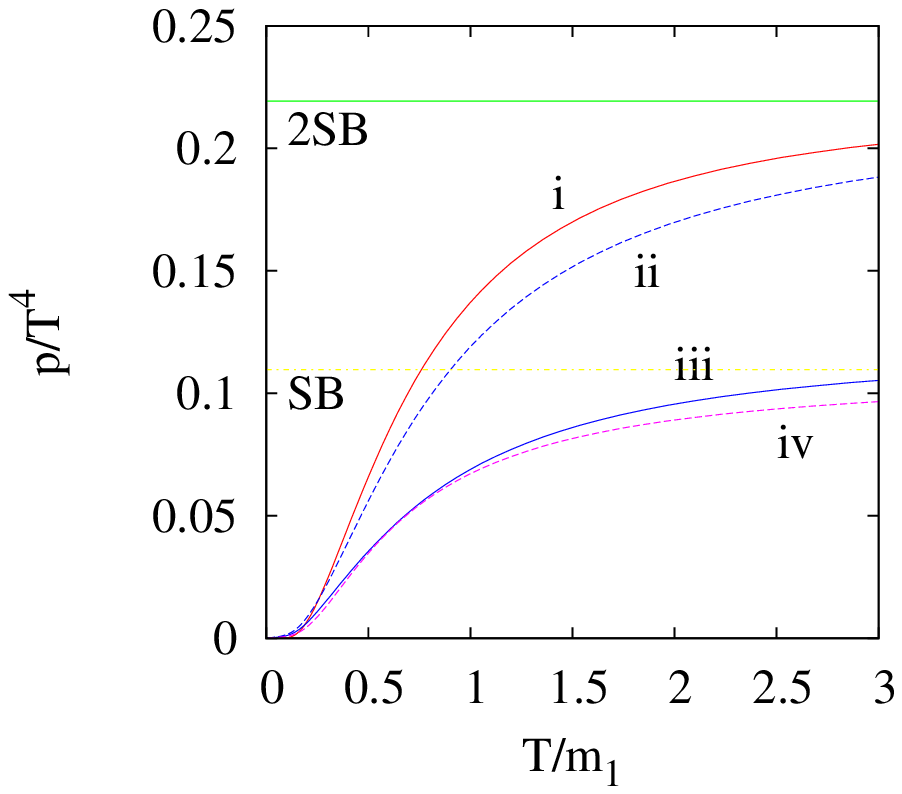}\\
  \hspace*{2em}\emph{a.)}\hspace*{20em}\emph{b.)}
  \caption{The $T^4$ normalized \emph{(a.)} energy density and
    \emph{(b.)} pressure of a system with spectral function consisting
    of two Lorentzian peaks with $m_1=1$, $m_2=2$ and common $\Gamma$
    width, cf. eq. \eqref{twolorrho}. The result for $\Gamma=0.2$ is
    in both figures denoted by \emph{iii.)}, the other curves are here
    for comparison. \emph{i.)} is the result of the two Dirac-delta
    case. \emph{ii.)} is the result of the independent resonance
    approximation. \emph{iv.)} is the result of one Lorentzian case
    with mass $m=1.2$ and width $\Gamma=0.2$. SB means Stefan-Boltzmann
    limit, 2SB is twice the Stefan-Boltzmann limit.}
  \label{fig:twolor_thermo}
\end{figure}

Although the number of dof describes well, what is going on, but we
still have to see, how it is realized in the physically observable
quantities. To this end we determined the energy density and pressure
for the case of small and large width. The result can be seen on
Fig. \ref{fig:twolor_thermo}. 
The exact result for $\Gamma=0.2$ is the curve labeled by
\emph{iii.)}. If we assumed that the two Lorentzians are independently
contribute to the thermodynamics, then we would obtain curve
\emph{ii.)}. For comparison we also show the results of the small
width case (ie. two Dirac-deltas) as curve \emph{i.)} and the result
of \emph{one Lorentzian} case with mass $m=1.2$ and width $\Gamma=0.2$
as curve \emph{iv.)}. It can be clearly seen that in the overlapping
Lorentzian case the independent resonance approximation fails. While
the results of the independent resonance approximation is close to the
two Dirac-delta peaks case, the exact result is closer to the one
Lorentzian case with appropriate parameters: ie. a one-dof
approximation. Therefore the results of the thermodynamical
calculation which provides physical observables supports the
conclusion coming from the analysis of the number of dof \eqref{np}:
for $\Gamma/\Delta m=0.2$ the system with two Lorentzian peaks
practically behave as a one-dof system.

\subsection{Thresholds}

We have seen in the previous subsection that two Breit-Wigner
resonances in general are combined in a non-linear way. This
phenomenon is in accordance with the scattering theory predictions
\cite{Svec}, where the S-matrix contributions of the resonances are
non-independent, but they satisfy in general a complicated unitarity
constraint equation. We should go further and study, what happens when
the spectral function can be represented by several resonances. Does
this nonlinear reduction of the number of dof take place also in this
case?

\begin{figure}[htbp]
  \centering
  \includegraphics[height=5cm]{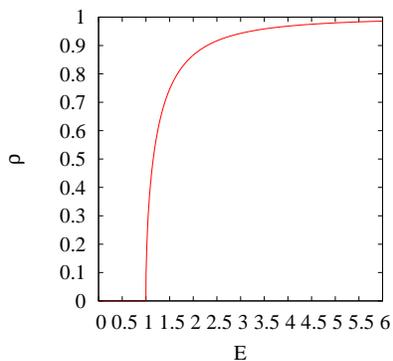}  
  \caption{The spectral function of a single threshold}
  \label{fig:threshold}
\end{figure}

The generic formulae for the energy density, pressure and number of
dof are shown in \eqref{eps}, \eqref{pressure} and \eqref{np}, they
can be evaluated for an arbitrary spectral function. Now let us choose
a simple threshold spectral function
\begin{equation}
  \rh(p) = \theta(p^2-M^2)\sqrt{1-\frac{M^2}{p^2}}.
\end{equation}
The spectral function is shown in Fig. \ref{fig:threshold}.
In principle one could represent this spectral function, with some
prescribed accuracy, with a number of massively overlapping Lorentzian
peaks. Since the overall normalization drops out from the formulae, we
do not even have to care about the correct normalization. The
threshold is, however, very far from a particle-like spectral
function: there is no peak, it does not vanish at infinity. We
actually do not even expect that it describes a physical system on its
own: rather it serves as a mathematical example for the strongly
correlated peaks.

\begin{figure}[htbp]
  \centering
  \includegraphics[height=5cm]{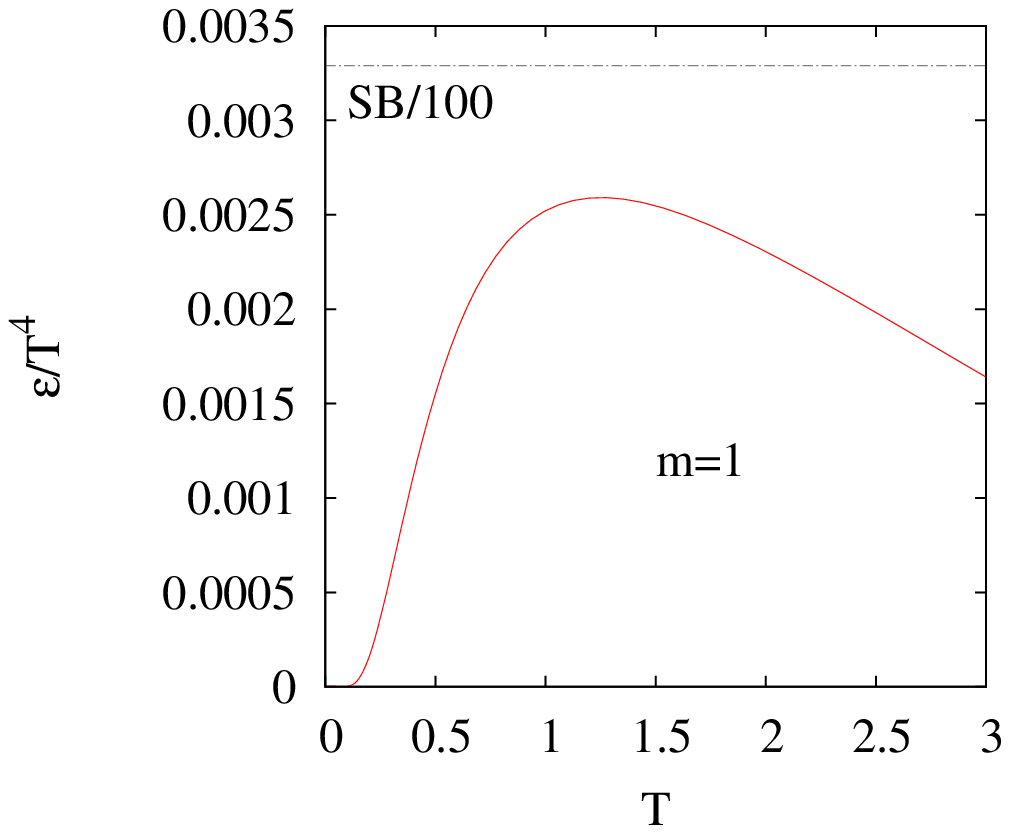}
  \includegraphics[height=5cm]{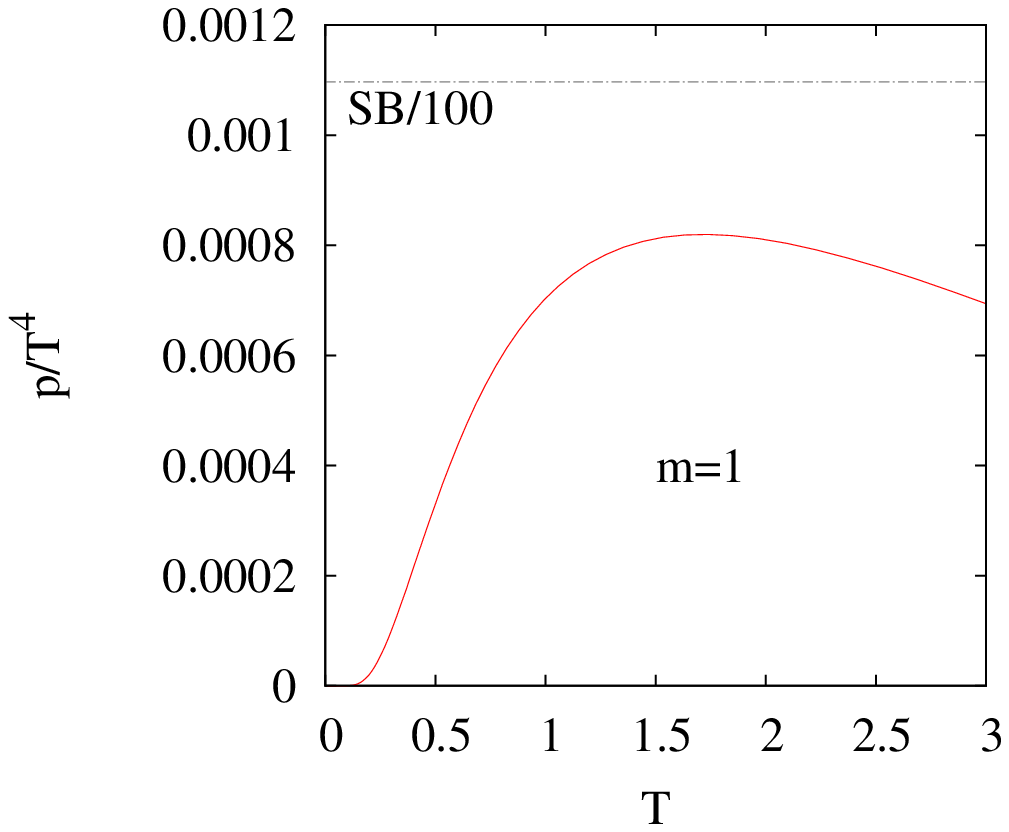}\\
  \hspace*{2em}\emph{a.)}\hspace*{20em}\emph{b.)}
  \caption{The energy density and pressure of the pure threshold
    spectral function. SB/100 means the hundredth of the
    Stefan-Boltzmann limit.}
  \label{fig:threholdthermo}
\end{figure}

When we evaluate the number of dof using \eqref{np} we obtain for
$m=1$ the value $N_{dof}=0.98$. However, when we use the formula for
the energy density and pressure we obtain the curves of
Fig. \ref{fig:threholdthermo}.
These curves are very far from the multi-dof expectations. The most
striking difference is in the amplitude: instead of having several
times the Stefan-Boltzmann limit, we are below the $0.01\times$ SB
limit. This further supports the hadron resonance gas approximation
where one neglects the contribution of the branch cuts to the
thermodynamics. We see that in this pathological case the number of
dof coming from \eqref{np} was also not informative, this seems to
work only for particle-like systems.

If we want to describe what is going on using the language of multiple
resonances, we can say that the contributions of the single resonances
interfere with each other so strongly, that their net contribution to
thermodynamical quantities is almost completely washed out. To
understand how is it possible we recall that the single resonance
contributions are complex (cf. \cite{Svec}), but their sum finally
must give a real number. So we have a large number of uncorrelated
complex phases, and study their sum: it is then plausible that finally
we obtain a near-zero net contribution.

We emphasize that this last example is non-physical. It serves for
providing a warning signal that it can be very misleading if one tries
to represent a spectral function with a large number of heavily
overlapping peaks.

\section{Conclusions}
\label{sec:conc}

In this paper we discussed the question of what kind of effective
model can be built on an experimentally determined spectral function.
The spectral function in general can consist of multiple finite width
peaks and non-particle-like continuum contributions. One procedure can
be to represent the spectral function with the sum of several
Breit-Wigner spectra, and assign a new dof for each of these
Breit-Wigner peaks which interact with each other in a way that the
correct widths come out (independent resonance
approximation). However, in this case each of the new degrees of
freedom at tree level introduces a new conserved quantity (the
corresponding particle number), which may alter the dynamical
behavior of the system. While for large separation this procedure is
known to work (cf. \cite{RD}), with massively overlapping peaks the
independence of these resonances is questionable.

A more conservative way is to find a representation which does not use
extra, non-physical dof. As a price we have to describe multiple peaks
with a single field. This can be done consistently, defining carefully
a non-local effective theory, and checking all the consistency
requirements. In this case the number of dof is not the number of the
representing fields (which is one), we can define it from the spectral
function.

Having a physical representation at hand, we can study how well the
independent resonance approximation works. We have found that for
infinite lifetime excitations (ie. a spectral function which consists
of discrete Dirac-delta peaks), the two methods provide identical
results. So, as we have expected, infinitely well separated peaks
indeed describe independent dof. But if the peaks are finite width
Breit-Wigner resonances and the peak separation becomes comparable
with the width, this nice agreement starts to fail. As we have shown,
the relevant parameter is $\Gamma/\Delta m$: a two-peak spectral
function with $\Gamma/\Delta m=0.2$ already behaves from the point of
view of thermodynamics as a one-dof system. An even more striking
example for the non-independence of the overlapping Breit-Wigner peaks
is the case of a mathematical construction of a threshold-like
spectral function. Although it could be represented with a sum of
multiple, overlapping Breit-Wigner resonances, the energy density of
this model is still about 100-times smaller than that of a
one-particle system.

We may conclude finally that a multi-dof representation of the
spectral functions can yield very misleading results. According to the
numerical evidences of this paper, the peaks of the spectral function
represent independent dof only in the case when the width of the peaks
is smaller than about $0.1\times$ the separation of the peaks. In
other cases one has to use a one-dof representation in order not to
change the original symmetries of the system, and obtain physically
correct results.

\section*{Acknowledgments}

The author is grateful to V. Koch for long and instructive
discussions, and for kind hospitality at BNL. The author thanks
D. Anchishkin, T.S. B\'{\i}r\'o, S.D. Katz, A. Patk\'os, P. Petreczky
and Zs. Sz\'ep for useful discussions. This work is supported by the
Hungarian Research Fund (OTKA) under contract No. K68108.

\end{document}